\documentclass[twocolumn,preprintnumbers,amsmath,amssymb,eqsecnum]{revtex4}

\usepackage{epsf}
\usepackage{graphicx}  % Include figure files
\usepackage{dcolumn}   % Align table columns on decimal point
\usepackage{bm}        % bold math

\begin{document}

\preprint{hep-th/0505133}

\title{Thermodynamics of Dual CFTs for Kerr-AdS Black Holes}

 \author{Rong-Gen Cai\footnote{e-mail address:
cairg@itp.ac.cn}}

\address{
  Institute of Theoretical Physics, Chinese
Academy of Sciences, P.O. Box 2735, Beijing 100080, China}

\author{Li-Ming Cao\footnote{e-mail address:
caolm@itp.ac.cn} and Da-Wei Pang\footnote{e-mail address:
pangdw@itp.ac.cn}}
\address{Institute of Theoretical Physics, Chinese
Academy of Sciences,
 P.O. Box 2735, Beijing 100080, China\\
  Graduate School of the Chinese Academy of Sciences, Beijing 100039, China}

\begin{abstract}
Recently Gibbons {\it et al.} in hep-th/0408217 defined a set of
conserved quantities for Kerr-AdS black holes with the maximal
number of rotation parameters in arbitrary dimension. This set of
conserved quantities is defined with respect to a frame which is
non-rotating at infinity. On the other hand, there is another set
of conserved quantities for Kerr-AdS black holes, defined by
Hawking {\it et al.} in hep-th/9811056, which is measured relative
to a frame rotating at infinity. Gibbons {\it et al.} explicitly
showed that the quantities defined by them satisfy the first law
of black hole thermodynamics, while those quantities defined by
Hawking {\it et al.} do not obey the first law. In this paper we
discuss thermodynamics of dual CFTs to the Kerr-AdS black holes by
mapping the bulk thermodynamic quantities to the boundary of the
AdS space. We find that thermodynamic quantities of dual CFTs
satisfy the first law of thermodynamics and Cardy-Verlinde formula
only when these thermodynamic quantities result from the set of
bulk quantities given by Hawking {\it et al.}. We discuss the
implication of our results.
\end{abstract}
\maketitle

%%========================section 1 ============================
\section{Introduction}

  Black holes in anti-de Sitter (AdS) space have been investigated
thoroughly in recent years due to the AdS/CFT
correspondence~\cite{AdS}. It was argued by Witten~\cite{witten}
that the thermodynamics of black holes in AdS spaces can be
identified with that of dual conformal field theory (CFT) residing
on the boundary of the AdS space. Therefore we can get some
insights into thermodynamic behavior and phase structure of some
strong coupling CFTs by studying corresponding thermodynamics of
black holes in AdS space.  One of interesting examples of black
holes in AdS space is Kerr-AdS black hole, a rotating black hole
in AdS space. According to the AdS/CFT correspondence, it was
argued by Hawking {\it et al.}~\cite{haw} that the thermodynamics
of Kerr-AdS black holes can be mapped to that of dual CFTs
residing on a rotating Einstein universe. In their paper, the
metric of Kerr-AdS black holes with a single rotation parameter in
arbitrary dimension was also given. Recently, the metric form of a
general Kerr-AdS black hole with the maximal number of rotation
parameters in arbitrary dimension has  been obtained
in~\cite{metric}.

To discuss thermodynamic properties of black holes in AdS space,
one has to first calculate the associated conserved charges with
the black hole configurations. On the other hand, one wants to
know corresponding thermodynamic quantities of dual CFTs via the
AdS/CFT correspondence. In the literature, there are different
methods to obtain conserved charges for spacetimes which are
asymptotically AdS, see, for example, references
\cite{ad,ht,am,by,bk,asd}. In general, all methods give consistent
results for static black holes in AdS space. For rotating black
holes in AdS space, however, some essential differences appear in
the literature. In particular, we would like to mention here that
Hawking {\it et al.} in \cite{haw} defined a set of conserved
quantities associated with Kerr-AdS black holes and this set of
quantities is measured with respect to a frame which is rotating
at infinity. On the other hand, very recently Gibbons {\it et
al.}~\cite{gib} also obtained a set of conserved quantities of
Kerr-AdS black holes, measured with a frame which is not-rotating
at infinity. And they further showed that the thermodynamic
quantities they obtained obey the first law of black hole
thermodynamics, while the set of conserved quantities derived by
Hawking {\it et al.} do not satisfy the first law. For other
earlier related studies on the Kerr-AdS black holes see
\cite{others} and references therein (The first reference in
\cite{others} is the first to discuss the first law for the four
dimensional Kerr-Newman-AdS black holes).

In this paper we will discuss thermodynamics of dual CFTs
associated with Kerr-AdS black holes and clarify the difference
between these two sets of conserved charges. By naively mapping
the thermodynamic quantities of Kerr-AdS black holes to those of
dual CFTs on the boundary of the AdS space, we find that the
resulting thermodynamic quantities from those defined by Hawking
{\it et al.} satisfy the first law of thermodynamics and the
entropy of CFTs can be expressed in terms of the Cardy-Verlinde
formula~\cite{Ver}, which is supposed to be an entropy formula of
strong coupling CFTs in arbitrary dimension. On the other hand,
the resulting thermodynamic quantities from those defined by
Gibbons {\it et al.} do not satisfy the first law of
thermodynamics, and the entropy cannot be written in the form of
the Cardy-Verlinde form, where a term associated with the pressure
of CFTs and volume of the boundary plays an essential role. Our
results indicate that in the AdS/CFT correspondence for the
Kerr-AdS black holes, the thermodynamic quantities of dual CFTs
associated with the Kerr-AdS black holes should be obtained by
mapping the bulk thermodynamic quantities of Kerr-AdS black holes
given by Hawking {\it et al.}, instead of those by Gibbons {\it et
al.}. This further supports that the dual CFTs to the Kerr-AdS
black holes reside on a rotating Einstein universe.

This paper is organized as follows. In the next section, we
briefly review two sets of thermodynamic quantities of four
dimensional Kerr-AdS black holes, given by Hawking {\it et
al}~\cite{haw} and by Gibbons {\it et al.}~\cite{gib}
respectively, and discuss the difference between them. In Sec.~III
we show that thermodynamic quantities of dual CFTs, by mapping
thermodynamic quantities in the definition due to Hawking {\it et
al.} of Kerr-AdS black holes with the maximal number of rotation
parameters satisfy the first law of thermodynamics. We then verify
in Sec.~IV that the entropy of dual CFTs can be expressed in terms
of the Cardy-Verlinde formula. We summarize our results in Sec.~V
with discussions on the implication of our results.

%%========================section 2=============================
\section{Conserved charges in Kerr-AdS spacetime}

The metric of a four dimensional  Kerr-AdS black hole can be
expressed as
\begin{eqnarray}
\label{2eq1}
 ds^{2} &=&
-\frac{\Delta}{\rho^{2}}(dt-\frac{a}{\Xi}\sin^{2}\theta d\phi)^{2}
+\frac{\rho^{2}dr^{2}}{\Delta}+\frac{\rho^{2}d\theta^{2}}{\Delta_{\theta}}
\nonumber \\
&&  +\frac{\Delta_{\theta}\sin
^{2}\theta}{\rho^{2}}(adt-\frac{r^{2}+a^{2}}{\Xi}d\phi)^{2},
\end{eqnarray}
where
\begin{equation}
\begin{array}{ll}
\Delta = (r^{2}+a^{2})(1+r^{2}l^{-2})-2mr,\ \Delta_{\theta}=
1-a^{2}l^{-2}\cos^{2}\theta, \\
\\
\rho^{2}= r^{2}+a^{2}\cos^{2}\theta, \ \ \ \Xi= 1-a^{2}l^{-2}.
\end{array}
\end{equation}
The integration constant $m$ is related to the mass of the black
hole and $a$ is the rotation parameter. The black hole horizon
$r_{+}$ is determined by the equation $\Delta(r)|_{r=r_+}=0$.

In \cite{gib} Gibbons {\it et al.} obtain the mass and angular
momentum associated with the Kerr-AdS black hole by using the
background subtraction approach
\begin{equation}
E=\frac{m}{\Xi^{2}},\ \ \ J=\frac{ma}{\Xi^{2}}.
\end{equation}
By using Hamiltonian approach, one can also obtain the same mass
and angular momentum~\cite{ht}. The Hawking temperature of the
black hole is easily derived by calculating the surface gravity on
the horizon
\begin{equation}
T=\frac{\kappa}{2\pi}=\frac{r_{+}(1+a^{2}l^{-2}+3r_{+}^{2}l^{-2}-a^{2}r_{+}^{-2})}{4\pi
(r_{+}^{2}+a^{2})},
\end{equation}
where $\kappa$ denotes the surface gravity. The entropy of the
black hole satisfies the area formula, one quarter of the horizon
area,
\begin{equation}
S=\frac{1}{4}A=\frac{\pi(r_{+}^{2}+a^{2})}{\Xi},
\end{equation}
where $A$ is the horizon area. Defining the angular velocity of
the black hole with respect to a frame which is non-rotating at
infinity~\cite{gib}
\begin{equation}
\label{2eq6}
 \Omega =
\frac{a(1+r_{+}^{2}l^{-2})}{r_{+}^{2}+a^{2}},
\end{equation}
Gibbons {\it et al.} \cite{gib} explicitly show  that these
thermodynamic quantities satisfy the first law of black hole
thermodynamics
\begin{equation}
dE=TdS+\Omega dJ.
\end{equation}
On the other hand, there exist other expressions of conserved
charges associated with the Kerr-AdS black hole. For example,
Hawking {\it et al.} give the expression
\begin{equation}
E'= \frac{m}{\Xi},
\end{equation}
for the mass, while still taking $J= ma/\Xi^2$ for the angular
momentum. Such a set of conserved charges can also be obtained by
using so-called boundary counterterm approach~\cite{bk}, or the
Weyl tensor approach suggested by Ashtekar {\it et
al.}~\cite{am,asd}. Further they define an angular velocity which
is measured relative to a frame rotating at infinity, by
\begin{equation}
\Omega'\equiv
-\frac{g_{t\phi}}{g_{\phi\phi}}\left|_{r=r_+}\right.=
\frac{a\Xi}{r_{+}^{2}+a^{2}}.
\end{equation}
This differs from $\Omega$  by
\begin{equation}
\Omega-\Omega'=\frac{a}{l^{2}}.
\end{equation}
Gibbons {\it et al.} verify that this set of thermodynamic
quantities do not satisfy the first law of black hole
thermodynamics
\begin{equation}
dE'\neq TdS+\Omega' dJ.
\end{equation}
In addition, they give a relation between $E$ and $E'$
\begin{equation}
E= E'+ Ja/l^2,
\end{equation}
which reflects a fact that the conserved charges $E$ and $E'$ are
associated with different Killing vectors; the former is related
to $(\partial_t+ al^{-2}\partial_{\phi})$, while the latter to
$\partial_t$.

%%=======================section 3==============================
\section{Thermodynamics of the dual CFT for Kerr-AdS black holes}

In the AdS/CFT correspondence, all thermodynamic quantities of
dual CFTs can be obtained by mapping bulk ones associated with
black holes in AdS space. In the previous section, we have seen
that there exist two sets of conserved charges for the Kerr-AdS
black holes: one is related to the Killing vector $\partial_t$
(corresponding to the case adopted by Hawking {\it et al.}), the
other is related to the Killing vector $\partial_t +
al^{-2}\partial_{\phi}$ (corresponding to the case adopted by
Gibbons {\it et al.}). It is then a natural question to ask which
set of conserved charges (thermodynamic quantities) of the
Kerr-AdS black holes should be mapped to obtain corresponding ones
of dual CFTs of the Kerr-AdS black holes.  According to the result
obtained in the previous section, it looks reasonable to map the
set of quantities given by Gibbons {\it et al.} and to derive
thermodynamic quantities of the boundary CFTs, because this set of
quantities satisfy the first law of black hole thermodynamics. It
turns out it is not correct and we should map the set of
quantities given by Hawking {\it et al.}, since only in this case,
resulting thermodynamic quantities of CFTs obey the first law of
thermodynamics. In this section we will show this with Kerr-AdS
black holes with the maximal number of rotation parameters in
arbitrary dimension.

 The metric of general Kerr-AdS black holes  with
the maximal number of rotation parameters in $n\ge 4$ dimensions
is~\cite{metric}
\begin{eqnarray}
\label{3eq1}
  ds^{2} &=& -W(1+r^{2}l^{-2})dt^{2}+\frac{2m}{U}(Wdt-
\sum\limits_{i=1}^{N}\frac{a_{i}\mu_{i}^{2}d\varphi_{i}}{\Xi_{i}})^{2}
\nonumber \\
 &+& \sum\limits_{i=1}^{N}\frac{r^{2}+a_{i}^{2}}{\Xi_{i}}\mu_{i}^{2}d\varphi_{i}^{2}
  +\frac{Udr^{2}}{V-2m}+\sum\limits_{i=1}^{N+\epsilon}\frac{r^{2}
+a_{i}^{2}}{\Xi_{i}}d\mu_{i}^{2} \nonumber
\\
&-& \frac{l^{-2}}{W(1+r^{2}l^{-2})}
(\sum\limits_{i=1}^{N+\epsilon}\frac{r^{2}+a_{i}^{2}}{\Xi_{i}}\mu_{i}d\mu_{i})^{2},
\end{eqnarray}
where
\begin{eqnarray}
&& W = \sum\limits_{i=1}^{N+\epsilon}\frac{\mu_{i}^{2}}{\Xi_{i}},\
\ \ U=
r^{\epsilon}\sum\limits_{i=1}^{N+\epsilon}\frac{\mu_{i}^{2}}{r^{2}
+a_{i}^{2}}\prod\limits_{j=1}^{N}(r^{2}+a_{j}^{2}), \nonumber \\
&& V =
r^{\epsilon-2}(1+r^{2}l^{-2})\prod\limits_{j=1}^{N}(r^{2}+a_{j}^{2}),\
 \nonumber \\
&& \Xi_{i}= 1-a_{i}^{2}l^{-2}.
\end{eqnarray}
Here $a_{i}$ stand for $N\equiv[(n-1)/2]$ independent rotation
parameters in $N$ orthogonal 2-planes. One has $n=2N+1$ when $n$
is odd and $n=2N+2$ when $n$ is even. $\epsilon= (n-1)$ mod $2$
and $n=2N+1+\epsilon$. In addition, $N$ azimuthal angles
$\phi_{i}$ and $(N+\epsilon)$ ``direction cosines" obeying
$\sum\limits_{i=1}^{N+\epsilon}\mu_{i}^{2}=1$ have  been also
introduced here.

The outer horizon $r_+$ of the black hole is determined by the
equation $V(r_+)-2m=0$. The Hawking temperature $T$ and the
entropy $S$ are easily obtained; they are given in~\cite{gib}
\begin{eqnarray}
&& n=odd: \nonumber \\
&& T=\frac{1}{2\pi}\left(
r_{+}(1+r_{+}^{2}l^{-2})\sum\limits_{i=1}^{N}\frac{1}{r_{+}^{2}+a_{i}^{2}}-\frac{1}{r_{+}}\right
 ), \nonumber
\\
&& S=\frac{\omega_{n-2}}{4r_{+}}\prod\limits_{i=1}^{N}\frac{r_{+}^{2}+a_{i}^{2}}{\Xi_{i}},
\end{eqnarray}
\begin{eqnarray}
&& n=even: \nonumber \\
&& T=\frac{1}{2\pi}\left
(r_{+}(1+r_{+}^{2}l^{-2})\sum\limits_{i=1}^{N}\frac{1}{r_{+}^{2}+a_{i}^{2}}
-\frac{1-r_{+}^{2}l^{-2}}{2r_{+}}\right), \nonumber \\
&&  S=\frac{\omega_{n-2}}
 {4}\prod\limits_{i=1}^{N}\frac{r_{+}^{2}+a_{i}^{2}}{\Xi_{i}},
\end{eqnarray}
where $\omega_{n-2}$ denotes the volume of the unit
$(n-2)$-sphere:
$$
\omega_{n-2}=\frac{2\pi^{(n-1)/2}}{\Gamma[(n-1)/2]}.
$$
Gibbons {\it et al.}~\cite{gib} find that by evaluating Komar
integrals, the angular momenta of the black holes associated with
the rotation parameters $a_i$ are given by
 \begin{equation}
 J_i= \frac{ma_i\omega_{n-2}}{4\pi \Xi_i\Pi_j\Xi_j}.
 \end{equation}
 And the energy of the Kerr-AdS black holes is
 \begin{eqnarray}
 n=odd:&& E= \frac{m\omega_{n-2}}{4\pi
 \Pi_j\Xi_j}\left(\sum^N_{i=1}\frac{1}{\Xi_i}-\frac{1}{2}\right),
 \\
 n=even:&& E=\frac{m\omega_{n-2}}{4\pi
 \Pi_j\Xi_j}\sum^N_{i=1}\frac{1}{\Xi_i}.
 \end{eqnarray}
The angular velocities of the black holes, measured with respect
to a frame which is non-rotating at infinity, are
\begin{equation}
\Omega_i=\frac{(1+r_+^2l^{-2})a_i}{r_+^2+a^2}.
\end{equation}
These thermodynamic quantities satisfy the first law of black hole
thermodynamics
\begin{equation}
\label{3eq9}
 dE=TdS +\sum^N_{i=1}\Omega_idJ_i.
\end{equation}
As the case in the four dimensional Kerr-AdS black holes, one can
obtain another mass (energy) of the black holes associated with
the Killing vector $\partial_t$
\begin{equation}
E'=\frac{m(n-2)\omega_{n-2}}{8\pi(\prod\limits_{i=1}^{N}\Xi_{i})},
\end{equation}
and angular velocities of the black holes
\begin{equation}
\Omega_i'=\frac{a_{i}\Xi_{i}}{r_{+}^{2}+a_{i}^2}.
\end{equation}
They are measured in a frame rotating at infinity. In such a
frame, other quantities keep unchanged. Naturally these quantities
do not obey the first law of black hole thermodynamics
\begin{equation}
\label{3eq12}
 dE' \ne TdS +\sum^N_{i=1}\Omega_i'dJ_i,
\end{equation}
once again.

Next we consider the thermodynamics of dual CFTs to the Kerr-AdS
black holes.  The dual CFTs reside on the boundary of the bulk
metric (\ref{3eq1}) in the spirit of AdS/CFT correspondence. The
boundary metric can be determined by using the bulk metric, up to
a conformal factor. Consider the four dimensional Kerr-AdS black
hole (\ref{2eq1}) as an example. Taking the conformal factor as
$R/r$, where $R$ is a constant with $R\gg l$, we have
\begin{eqnarray}
\label{3eq13}
 ds^2_{CFT} &=& \lim_{r\to \infty}\frac{R^2}{r^2} ds^2_4
    = \frac{R^2}{l^2}( -dt^2 +\frac{2 a
    \sin^2\theta}{\Xi}dtd\phi
    \nonumber \\
    && +
    \frac{l^2}{\triangle_{\theta}}d\theta^2 +\frac{l^2
    \sin^2\theta}{\Xi}d\phi^2) \nonumber \\
 &=& \frac{R^2}{l^2}( -\frac{\triangle_{\theta}}{\Xi}dt^2
   + \frac{l^2}{\triangle_{\theta}}d\theta^2
    \nonumber \\
    && + \frac{l^2\sin^2\theta}{\Xi}(d\phi+al^{-2}dt)^2).
\end{eqnarray}
Rescaling the time coordinate as
\begin{equation}
\label{rescaling}
 Rdt/l =d\tau
 \end{equation}
  we reach
\begin{eqnarray}
\label{boundary}
 ds^2_{CFT} &=&-\frac{\triangle_{\theta}}{\Xi}d\tau^2
   + \frac{R^2}{\triangle_{\theta}}d\theta^2
     \nonumber \\
   && + \frac{R^2\sin^2\theta}{\Xi}(d\phi+al^{-1}
   R^{-1}d\tau)^2 \nonumber \\
   &=&-d\tau^2 +\frac{2 a R
    \sin^2\theta}{l\Xi}d\tau d\phi
    \nonumber \\
    && +
    \frac{R^2}{\triangle_{\theta}}d\theta^2 +\frac{R^2
    \sin^2\theta}{\Xi}d\phi^2.
    \end{eqnarray}
Note that the boundary metric (\ref{boundary}) describes a
(2+1)-dimensional rotating Einstein universe with scale factor
$R$. And the spatial volume is $V=4 \pi R^2/\Xi $. Furthermore, we
can see that the angular velocities for the rotating Einstein
universe are
$$\Omega_{\infty} = - al^{-2}, \ \ \ \Omega_{\infty}=-a
(lR)^{-1},$$ in the metric (\ref{3eq13}) and (\ref{boundary}),
respectively.  For higher dimensional case, the situation is
similar to the four dimensional case. That is, the boundary metric
where the dual CFTs reside in can be a rotating Einstein universe
with  scale factor $R$. We now can obtain the corresponding
thermodynamic quantities of dual CFTs by naively mapping
quantities associated with the bulk to those of the dual CFTs on
the boundary (\ref{boundary}). Due to the rescaling relation
(\ref{rescaling}), the expressions for the energy, temperature and
angular velocity of the dual CFTs have the forms~\cite{SV} (see
also \cite{Cai,Birm,Jing,Muck})
\begin{eqnarray}
\label{3eq14}
 && E_{CFT}(E'_{CFT})=\frac{l}{R}E(E'),\ \ \
T_{CFT}=\frac{l}{R}T,  \nonumber \\
&& \Omega_{CFT} (\Omega'_{CFT}) = \frac{l}{R}\Omega (\Omega').
\end{eqnarray}
In the mapping process, the angular momentum $J$ and entropy $S$
remain unchanged:
 \begin{equation}
 \label{entropy}
  J_{CFT}=J, \ \ \ S_{CFT}=S,
  \end{equation}
since entropy stands for the degrees of freedom of system and the
angular momentum is conserved charge associated with the Killing
vector $\partial/\partial\phi$. Note that here our rescaling
relations for the angular velocity (\ref{3eq14}) and angular
momentum (\ref{entropy}) are different from those in \cite{Jing}.
We think our rescaling relations are correct. In addition, let us
mention that since we are considering $(n-1)$-dimensional CFTs
which reside on the boundary with volume $V$, the pressure of the
CFTs has a simple relation to the energy $E$ and volume:
\begin{equation}
P= \frac{E}{(n-2)V}.
 \end{equation}

Now we  consider the mapping of the set of thermodynamic
quantities in the prescription given by Hawking {\it et al.}.

i) when $n=odd$, one has  $\epsilon=0$, and
\begin{eqnarray}
&& E_{CFT}'= \frac{m l(n-2)\omega_{n-2}}{8\pi R(\Pi_i\Xi_{i})},\ \
\
J_{iCFT}=\frac{ma_{i}\omega_{n-2}}{4\pi \Xi_{i}(\Pi_j\Xi_{j})}, \nonumber \\
 && T_{CFT} = \frac{l}{2\pi R}\left (r_{+}(1+r_{+}^{2}l^{-2})\sum\limits_{i=1}^{N}
\frac{1}{r_{+}^{2}+a_{i}^{2}}-\frac{1}{r_{+}}\right ), \nonumber \\
&& S
=\frac{\omega_{n-2}}{4r_{+}}\prod\limits_{i=1}^{N}\frac{r_{+}^{2}+a_{i}^{2}}{\Xi_{i}},
\ \ \
 \Omega_{iCFT} = \frac{la_{i}\Xi_{i}}{R(r_{+}^{2}+a_{i}^2)}.
\end{eqnarray}
The  spatial volume of the boundary
\begin{equation}
V=\frac{\omega_{n-2}R^{2N-1}}{\Pi_{i}\Xi_{i}},
\end{equation}
and then the pressure of the CFTs
\begin{equation}
P'=\frac{E_{CFT}'}{(n-2)V}.
\end{equation}

ii) when $n=even$, one has  $\epsilon=1$, and
\begin{eqnarray}
&& E_{CFT}' = \frac{ml(n-2)\omega_{n-2}}{8\pi R(\Pi_i\Xi_{i})},\ \
\ J_{iCFT}=\frac{ma_{i}\omega_{n-2}}{4\pi
\Xi_{i}(\Pi_j\Xi_{j})}, \nonumber \\
&&  T_{CFT} =\frac{l}{2\pi R}\left
(r_{+}(1+r_{+}^{2}l^{-2})\sum\limits_{i=1}^{N}\frac{1}{r_{+}^{2}+a_{i}^{2}}
-\frac{1-r_{+}^{2}l^{-2}}{2r_{+}}\right),\nonumber \\
&&
 S=\frac{\omega_{n-2}}{4}\prod\limits_{i=1}^{N}\frac{r_{+}^{2}+a_{i}^{2}}{\Xi_{i}},
\ \ \
 \Omega_{iCFT}' =\frac{la_{i}\Xi_{i}}{R(r_{+}^{2}+a_{i}^2)}.
\end{eqnarray}
In this case, the volume $V$ and pressure of the CFTs are
\begin{equation}
V=\frac{\omega_{n-2}R^{2N}}{\Pi_i\Xi_{i}},\ \ \
P'=\frac{E_{CFT}'}{(n-2)V},
\end{equation}
respectively. We find that these thermodynamic quantities satisfy
the first law of thermodynamics
\begin{equation}
\label{firstlaw}
dE_{CFT}'=T_{CFT}dS+\sum_{i=1}^{N}\Omega_{iCFT}'dJ_{i}-P'dV.
\end{equation}
Its validity from $n=4$ up to $n=11$ has been checked by using
{\it Mathematics}. Note that here  the scale factor $R$ in
(\ref{firstlaw}) is variable. When considering $R$ as a constant
and setting $R=l$, we can see from (\ref{3eq14}) and
(\ref{entropy}) that all quantities for the dual CFTs reduce to
corresponding ones for Kerr-AdS black holes. Indeed, in this case,
the first law (\ref{firstlaw}) for the dual CFTs  is found to
degenerate to the first law (\ref{3eq9}) for the black holes.

 On the other hand, if we map the set of
thermodynamic quantities given by Gibbons {\it et al.} according
to the relation (\ref{3eq14}) and (\ref{entropy}), resulting
thermodynamic quantities of dual CFTs do not satisfy the first law
of thermodynamics
\begin{equation}
dE_{CFT} \ne T_{CFT} dS +\sum^N_{i=1} \Omega_{iCFT} dJ_{i}-PdV.
\end{equation}
Therefore we conclude that the thermodynamic quantities of dual
CFTs to the Kerr-AdS black holes satisfy the first law of
thermodynamics if we map the set of bulk thermodynamic quantities
given in the prescription of Hawking {\it et al.}, instead they do
not if we map the set of bulk thermodynamic quantities given by
Gibbons {\it et al.}. The results can explained as follows. From
the metric (\ref{boundary}), we can see that the dual CFTs reside
on a rotating Einstein universe, where the timelike Killing vector
is $\partial/\partial \tau$. Therefore the energy of the CFTS
should be measured respect to this Killing vector. In other words,
all physical quantities should be measured in the rotating frame.
Our results also further confirm the argument by Hawking et al.
\cite{haw} that the dual CFTs of Kerr-AdS black holes resides in a
rotating Einstein universe. In addition, from the result given by
Gibbons {\it et al.} it seems to tell us that black hole
thermodynamics should be measured in a frame which is static at
infinity, or with respect to static observers at infinity.

%%=======================section 4==============================
\section{Cardy-Verlinde formula for Kerr-AdS black holes}

In order to further verify that thermodynamic quantities of dual
CFTs to the Kerr-AdS black holes should be obtained by mapping the
set of bulk thermodynamic quantities given in the prescription of
Hawking {\it et al.}, in this section, we will show that only the
thermodynamic quantities of dual CFTs by mapping the set of bulk
thermodynamic quantities given in the prescription of Hawking {\it
et al.} obey the Cardy-Verlinde formula~\cite{Ver}, and resulting
thermodynamic quantities from the set of bulk quantities given by
Gibbons {\it et al.} do not satisfy the Cardy-Verlinde formula.

There is a well-known entropy formula for a $(1+1)$-dimensional
conformal field theory, namely, the Cardy formula~\cite{Cardy}. In
an elegant paper~\cite{Ver}, in the spirit of AdS/CFT
correspondence, Verlinde argued that there is a similar entropy
formula  for CFTs in higher dimensions. The formula ``derived" by
Verlinde is called Cardy-Verlinde formula in the literature.
Indeed this formula has been checked to hold for various CFTs with
AdS gravity duals, such as Schwarzschild-AdS black
holes~\cite{Ver}, Kerr-AdS black holes~\cite{KPS}, Hyperbolic and
charged black holes~\cite{Cai}, Taub-Bolt-AdS
instanton~\cite{Birm}, Kerr-Newmann-AdS black holes~\cite{Jing}
and so on (see also \cite{Muck} and references therein). In
addition, it has been found that entropies of black hole
horizons~\cite{Cai2} and cosmological horizons~\cite{Cai3} in
asymptotically dS spaces can also be expressed in terms of the
Cardy-Verlinde formula.  The Cary-Verlinde formula is supposed to
be an entropy formula for CFTs in arbitrary dimension.

 Consider a CFT
living in $(n-1)$-dimensional spacetime described by the metric
\begin{equation}
\label{4eq1}
 ds^{2}=-dt^{2}+R^{2}d\Omega_{n-2}^{2},
\end{equation}
where $R$ is the radius of an $(n-2)$-dimensional sphere. The
Cardy-Verlinde formula can be expressed as
\begin{equation}
S=\frac{2\pi R}{n-2}\sqrt{E_{c}(2E-E_{c})},
\end{equation}
where $E_{c}$ denotes the Casimir energy, non-extensive part of
the energy $E$ of CFT,  defined as
\begin{equation}
\label{4eq3}
E_{c}\equiv(n-2)(E+PV-TS)=(n-1)E-(n-2)TS.
\end{equation}
When some chemical potentials appear in some CFTs, more terms
associated with these chemical potentials should be included to
the definition (\ref{4eq3}). For example, for the case of rotating
black holes, $E_c$ in (\ref{4eq3}) should be modified  to
$E_{c}\equiv(n-2)(E+PV-TS-\Omega J)$.

With the thermodynamic quantities of dual CFTs given in the
previous section, we can see that the Cardy-Verlinde formula holds
if we map the set of bulk thermodynamic quantities given in the
prescription of Hawking {\it et al.}. Here the Casimir energy
$E_{c}$ is defined as
\begin{eqnarray}
\label{4eq4}
E_{c} &\equiv & (n-2)(E_{CFT}'+P'V-T_{CFT}S-\sum\limits_{i=1}^{N}\Omega_{iCFT}'J_{i}) \nonumber \\
 &=&(n-1)E_{CFT}-(n-2)T_{CFT}S \nonumber \\
&& -(n-2)\sum\limits_{i=1}^{N}\Omega_{iCFT}'J_{i}).
\end{eqnarray}
A straightforward calculation gives

 i) when $n=odd$,
\begin{eqnarray}
 && E_{c}=\frac{(n-2)l\omega_{n-2}}{8\pi
Rr_{+}^{2}}\prod\limits_{i=1}^{N}\frac{r_{+}^{2}+a_{i}^2}{\Xi_{i}},
\nonumber
\\
&& 2E_{CFT}'-E_{c}=\frac{(n-2)\omega_{n-2}}{8\pi
lR}\prod\limits_{i=1}^{N}\frac{r_{+}^{2}+a_{i}^2}{\Xi_{i}},
\nonumber
\\
&& \frac{2\pi
R}{n-2}\sqrt{E_{c}(2E_{CFT}'-E_{c})}=\frac{\omega_{n-2}}{4r_{+}}\prod\limits_{i=1}^{N}
\frac{r_{+}^{2}+a_{i}^{2}}{\Xi_{i}}\nonumber \\
&&~~~~~~~~~~~~~~~~~~~~~~~~~~~~~~~~~=S.
\end{eqnarray}

ii) when $n=even$,
\begin{eqnarray}
&& E_{c}=\frac{(n-2)l\omega_{n-2}}{8\pi
Rr_{+}}\prod\limits_{i=1}^{N}\frac{r_{+}^{2}+a_{i}^2}{\Xi_{i}},
\nonumber
\\
&& 2E_{CFT}'-E_{c}=\frac{(n-2)\omega_{n-2}r_{+}}{8\pi
lR}\prod\limits_{i=1}^{N}\frac{r_{+}^{2}+a_{i}^2}{\Xi_{i}},\nonumber
\\
&&  \frac{2\pi
R}{n-2}\sqrt{E_{c}(2E_{CFT}'-E_{c})}=\frac{\omega_{n-2}}{4}\prod\limits_{i=1}^{N}\frac{r_{+}^{2}
+a_{i}^{2}}{\Xi_{i}}\nonumber \\
&&~~~~~~~~~~~~~~~~~~~~~~~~~~~~~~~~~=S.
\end{eqnarray}
Thus we have explicitly checked that the Cardy-Verlinde formula
holds for the dual CFTs to the Kerr-AdS black holes in arbitrary
dimension, if resulting thermodynamic quantities are obtained by
mapping the set of bulk thermodynamic quantities in the
prescription of Hawking {\it et al.}. On the other hand, if we
replace $E_{CFT}'$ by $E_{CFT}$, $\Omega_{iCFT}'$ by
$\Omega_{iCFT}$, and $P'$ by $P$ in (\ref{4eq4}), it is easy to
check that the entropy of dual CFTs cannot be recast into the
Cardy-Verlinde form
\begin{equation}
S \ne \frac{2\pi R}{n-2}\sqrt{E_c(2E_{CFT}-E_c)}.
\end{equation}
Here it might be worth mentioning again that the dual CFTs of the
Kerr-AdS black holes do not reside in the spacetime (\ref{4eq1}),
a static sphere, but on a rotating Einstein universe (sphere) like
(\ref{boundary}).

%%==========================section 5=========================
\section{Conclusion and discussion }

For the Kerr-AdS  black holes in arbitrary dimension, there exist
two sets of thermodynamic quantities: one is given by Hawking {\it
et al.}~\cite{haw}, and is measured in a frame rotating at
infinity; the other is given by Gibbons {\it et al.}~\cite{gib},
which is measured in a frame which is non-rotating at infinity.
And very recently Gibbons {\it et al.} have shown that only the
set of thermodynamic quantities given by them satisfy the first
law of black hole thermodynamics.  In this paper we have
investigated the thermodynamics of dual CFTs to the Kerr-AdS black
holes with the maximal number of rotation parameters.  We have
found that thermodynamic quantities of dual CFTs resulting from
the set of bulk thermodynamic quantities given in the prescription
given by Hawking {\it et al.} satisfy the first law of
thermodynamics and Cardy-Verlinde formula, which is supposed to be
an entropy formula of CFTs in higher dimensions, instead the
thermodynamic quantities of dual CFTs obtained by mapping the set
of bulk thermodynamic quantities given by Gibbons {\it et al.} do
not obey the first law of thermodynamics and cannot be rewritten
in terms of the Cardy-Verlinde formula.

Our results further indicate that the dual CFTs to the Kerr-AdS
black holes reside in a boundary which is a rotating Einstein
universe. Thermodynamic quantities of dual CFTs should be obtained
by mapping the set of bulk thermodynamic quantities given in the
prescription of Hawking {\it et al.}, although this set of bulk
thermodynamic quantities do not obey the first law of black hole
thermodynamics. On the other hand, from the result obtained by
Gibbons {\it et al.} one can see that black hole thermodynamics
should seemingly be measured with respect to static observers at
infinity.

\section*{Acknowledgements}
One of authors (DWP) thanks Hong-Sheng Zhang, Zong-Kuan Guo, Qi
Guo, Jian-Huang She for useful discussions and kind help. This
work was supported by grants from NSFC, China (No. 13325525 and
No. 90403029), and a grant from the Ministry of Science and
Technology of China (No. TG1999075401).


\begin{references}

\bibitem{AdS}J.~Maldacena,
%``The large N limit of superconformal field theories and supergravity,''
Adv.\ Theor.\ Math.\ Phys.\  {\bf 2}, 231 (1998) [Int.\ J.\ Theor.\
Phys.\  {\bf 38}, 1113 (1998)] [arXiv:hep-th/9711200];
%%CITATION = HEP-TH 9711200;%%
 S.~S.~Gubser, I.~R.~Klebanov and A.~M.~Polyakov,
%``Gauge theory correlators from non-critical string theory,''
Phys.\ Lett.\ B {\bf 428}, 105 (1998) [arXiv:hep-th/9802109];
%%CITATION = HEP-TH 9802109;%%
 E.~Witten,
%``Anti-de Sitter space and holography,''
Adv.\ Theor.\ Math.\ Phys.\  {\bf 2}, 253 (1998)
[arXiv:hep-th/9802150].
%%CITATION = HEP-TH 9802150;%%

\bibitem{witten}E.~Witten,
 %``Anti-de Sitter space, thermal phase transition, and confinement in  gauge
%theories,''
Adv.\ Theor.\ Math.\ Phys.\  {\bf 2}, 505 (1998)
[arXiv:hep-th/9803131].
%%CITATION = HEP-TH 9803131;%%

\bibitem{haw}S.~W.~Hawking, C.~J.~Hunter and M.~M.~Taylor-Robinson,
%"Rotation and the AdS/CFT correspondence,"
Phys.\ Rev.\ D{\bf 59},064005 (1999) [arXiv:hep-th/9811056].
%%CITATION = HEP-TH 9811056;%%

\bibitem{metric}G.~W.~Gibbons, H.~Lu, D.~N.~Page and C.~N.~Pope,
  %``The general Kerr-de Sitter metrics in all dimensions,''
  J.\ Geom.\ Phys.\  {\bf 53}, 49 (2005)
  [arXiv:hep-th/0404008];
  %%CITATION = HEP-TH 0404008;%%
G.~W.~Gibbons, H.~Lu, D.~N.~Page and C.~N.~Pope,
  %``Rotating black holes in higher dimensions with a cosmological constant,''
  Phys.\ Rev.\ Lett.\  {\bf 93}, 171102 (2004)
  [arXiv:hep-th/0409155].
  %%CITATION = HEP-TH 0409155;%%


\bibitem{ad}L.~F.~Abbott and S.~Deser,
%``Stability Of Gravity With A Cosmological Constant,''
Nucl.\ Phys.\ B {\bf 195}, 76 (1982).
%%CITATION = NUPHA,B195,76;%%

\bibitem{ht}M. Henneaux and C. Teitelboim,
%{\it Asymptotically anti-de Sitter spaces},
Commun.\ Math.\ Phys.\ {\bf 98}, 391 (1985).

\bibitem{am}A. Ashtekar and A.Magnon,
%{\it Asymptotically anti-de Sitter space-times},
Class.\ Quant.\ Grav.\ {\bf 1}, L39 (1984).

\bibitem{by}J.D. Brown and J.W. York,
%{\it Quasilocal energy and conserved charges derived from the gravitational action},
Phys.\ Rev.\ {\bf D47}, 1407 (1993).
%%CITATION = PHRVA,D47,1407;%%

\bibitem{bk}V. Balasubramanian and P. Kraus,
%{\it A stress tensor for anti-de Sitter gravity},
Commun.\ Math.\ Phys.\ {\bf 208}, 413 (1999),
[arXiv:hep-th/9902121].

\bibitem{asd}A. Ashtekar and S. Das,
%{\it Asymptotically anti-de Sitter space-times: Conserved quantities},
Class.\ Quant.\ Grav.\ {\bf 17}, L17 (2000), [arXiv:hep-th/9911230].


\bibitem{gib}G.W. Gibbons, M.J. Perry and C.N. Pope,
%{\it The first law of thermodynamics for Kerr-anti-de Sitter black holes},
[arXiv:hep-th/0408217], to appear in Class. Quantum Grav.
%%CITATION = HEP-TH 0408217;%%



\bibitem{others}
  M.~M.~Caldarelli, G.~Cognola and D.~Klemm,
  %``Thermodynamics of Kerr-Newman-AdS black holes and conformal field
  %theories,''
  Class.\ Quant.\ Grav.\  {\bf 17}, 399 (2000)
  [arXiv:hep-th/9908022];
  %%CITATION = HEP-TH 9908022;%%
  A.~M.~Awad and C.~V.~Johnson,
  %``Holographic stress tensors for Kerr-AdS black holes,''
  Phys.\ Rev.\ D {\bf 61}, 084025 (2000)
  [arXiv:hep-th/9910040];
  %%CITATION = HEP-TH 9910040;%%
  S.~Das and R.~B.~Mann,
  %``Conserved quantities in Kerr-anti-de Sitter spacetimes in various
  %dimensions,''
  JHEP {\bf 0008}, 033 (2000)
  [arXiv:hep-th/0008028];
  %%CITATION = HEP-TH 0008028;%%
A.~M.~Awad and C.~V.~Johnson,
  %``Higher dimensional Kerr-AdS black holes and the AdS/CFT correspondence,''
  Phys.\ Rev.\ D {\bf 63}, 124023 (2001)
  [arXiv:hep-th/0008211];
  %%CITATION = HEP-TH 0008211;%%
M.~H.~Dehghani and R.~B.~Mann,
  %``Quasilocal thermodynamics of Kerr and Kerr-anti-de Sitter spacetimes  and
  %the AdS/CFT correspondence,''
  Phys.\ Rev.\ D {\bf 64}, 044003 (2001)
  [arXiv:hep-th/0102001];
  %%CITATION = HEP-TH 0102001;%%
  N.~Deruelle and J.~Katz,
  %``On the mass of a Kerr-anti-de Sitter spacetime in D dimensions,''
  Class.\ Quant.\ Grav.\  {\bf 22}, 421 (2005)
  [arXiv:gr-qc/0410135].
  %%CITATION = GR-QC 0410135;%%


\bibitem{Ver}
E.~Verlinde,
%``On the holographic principle in a radiation dominated universe,''
arXiv:hep-th/0008140.
%%CITATION = HEP-TH 0008140;%%

\bibitem{SV}I.~Savonije and E.~Verlinde,
  %``CFT and entropy on the brane,''
  Phys.\ Lett.\ B {\bf 507}, 305 (2001)
  [arXiv:hep-th/0102042].
  %%CITATION = HEP-TH 0102042;%%



\bibitem{Cardy}J.~L.~Cardy,
%``Operator Content Of Two-Dimensional Conformally Invariant Theories,''
Nucl.\ Phys.\ B {\bf 270}, 186 (1986).
%%CITATION = NUPHA,B270,186;%%


\bibitem{KPS}
 D.~Klemm, A.~C.~Petkou and G.~Siopsis,
%``Entropy bounds, monotonicity properties and scaling in CFTs,''
Nucl.\ Phys.\ B {\bf 601}, 380 (2001) [arXiv:hep-th/0101076].
%%CITATION = HEP-TH 0101076;%%

\bibitem{Cai}
 R.~G.~Cai,
%``The Cardy-Verlinde formula and AdS black holes,''
Phys.\ Rev.\ D {\bf 63}, 124018 (2001) [arXiv:hep-th/0102113];
%%CITATION = HEP-TH 0102113;%%
R.~G.~Cai, Y.~S.~Myung and N.~Ohta,
%``Bekenstein bound, holography and brane cosmology in charged black hole
%background,''
Class.\ Quant.\ Grav.\  {\bf 18}, 5429 (2001)
[arXiv:hep-th/0105070];
%%CITATION = HEP-TH 0105070;%%
R.~G.~Cai and A.~z.~Wang,
%``Thermodynamics and stability of hyperbolic charged black holes,''
Phys.\ Rev.\ D {\bf 70}, 064013 (2004) [arXiv:hep-th/0406057].
%%CITATION = HEP-TH 0406057;%%

\bibitem{Birm}
 D.~Birmingham and S.~Mokhtari,
%``The Cardy-Verlinde formula and Taub-Bolt-AdS spacetimes,''
Phys.\ Lett.\ B {\bf 508}, 365 (2001) [arXiv:hep-th/0103108].
%%CITATION = HEP-TH 0103108;%%
\bibitem{Jing}
J.~l.~Jing,
%``Cardy-Verlinde formula and Kerr-Newman-AdS(4) and two rotation  parameters
%Kerr-AdS(5) black holes,''
Phys.\ Rev.\ D {\bf 66}, 024002 (2002) [arXiv:hep-th/0201247].
%%CITATION = HEP-TH 0201247;%%

\bibitem{Muck}A.~K.~Biswas and S.~Mukherji,
  %``Holography and stiff-matter on the brane,''
  JHEP {\bf 0103}, 046 (2001)
  [arXiv:hep-th/0102138];
  %%CITATION = HEP-TH 0102138;%%
R.~G.~Cai and Y.~Z.~Zhang,
  %``Holography and brane cosmology in domain wall backgrounds,''
  Phys.\ Rev.\ D {\bf 64}, 104015 (2001)
  [arXiv:hep-th/0105214];
  %%CITATION = HEP-TH 0105214;%%
L.~Cappiello and W.~Muck,
%``On the phase transition of conformal field theories with holographic
%duals,''
Phys.\ Lett.\ B {\bf 522}, 139 (2001) [arXiv:hep-th/0107238];
%%CITATION = HEP-TH 0107238;%%
S.~Nojiri, S.~D.~Odintsov and S.~Ogushi,
%``Holographic entropy and brane FRW dynamics from AdS black hole in d5  higher
%derivative gravity,''
Int.\ J.\ Mod.\ Phys.\ A {\bf 16}, 5085 (2001)
[arXiv:hep-th/0105117];
%%CITATION = HEP-TH 0105117;%%
S.~Nojiri, S.~D.~Odintsov and S.~Ogushi,
%``Cosmological and black hole brane world universes in higher derivative
%gravity,''
Phys.\ Rev.\ D {\bf 65}, 023521 (2002) [arXiv:hep-th/0108172];
%%CITATION = HEP-TH 0108172;%%
S.~Nojiri, S.~D.~Odintsov and S.~Ogushi,
%``Friedmann-Robertson-Walker brane cosmological equations from the
%five-dimensional bulk (A)dS black hole,''
Int.\ J.\ Mod.\ Phys.\ A {\bf 17}, 4809 (2002)
[arXiv:hep-th/0205187];
%%CITATION = HEP-TH 0205187;%%
D.~Astefanesei, R.~Mann and E.~Radu,
%``Reissner-Nordstroem-de Sitter black hole, planar coordinates and dS/CFT,''
JHEP {\bf 0401}, 029 (2004) [arXiv:hep-th/0310273];
%%CITATION = HEP-TH 0310273;%%
R.~G.~Cai, D.~W.~Pang and A.~Wang,
  %``Born-Infeld black holes in (A)dS spaces,''
  Phys.\ Rev.\ D {\bf 70}, 124034 (2004)
  [arXiv:hep-th/0410158].
  %%CITATION = HEP-TH 0410158;%%



\bibitem{Cai2}R.~G.~Cai,
%``Cardy-Verlinde formula and thermodynamics of black holes in de Sitter
%spaces,''
Nucl.\ Phys.\ B {\bf 628}, 375 (2002) [arXiv:hep-th/0112253].
%%CITATION = HEP-TH 0112253;%%

\bibitem{Cai3}R.~G.~Cai,
%``Cardy-Verlinde formula and asymptotically de Sitter spaces,''
Phys.\ Lett.\ B {\bf 525}, 331 (2002) [arXiv:hep-th/0111093].
%%CITATION = HEP-TH 0111093;%%



\end{references}
\end{document}